# Automatic answering of scientific questions using the FACTS-V1 framework: New methods in research to increase efficiency through the use of AI

Dr. Stefan Pietrusky[1]

[1]Down Church Studios, Hintergasse 46a, 67150 Niederkirchen, Germany
[1]Heidelberg University of Education
[1]`pietrusky@downchurch.studio`

December 1, 2024

## Abstract

The use of artificial intelligence (AI) offers various possibilities to expand and support educational research. Specifically, the implementation of AI can be used to develop new frameworks to establish new research tools that accelerate and meaningfully expand the efficiency of data evaluation and interpretation [1]. This article presents the prototype of the FACTS-V1 (Filtering and Analysis of Content in Textual Sources) framework. With the help of the application, numerous scientific papers can be automatically extracted, analyzed and interpreted from open access document servers without having to rely on proprietary applications and their limitations. The FACTS-V1 prototype consists of three building blocks. The first part deals with the extraction of texts, the second with filtering and interpretation, and the last with the actual statistical evaluation (topic modeling) using an interactive overview. The aim of the framework is to provide recommendations for future scientific questions based on existing data. The functionality is illustrated by asking how the use of AI will change the education sector. The data used to answer the question comes from 82 scientific papers on the topic of AI from 2024. The papers are publicly available on the peDOCS document server of the Leibniz Institute for Educational Research and Educational Information.

## 1 Introduction

The possible uses of artificial intelligence (AI) are diverse. How the use of AI will change education in schools and universities is the subject of current discussions in educational research [2] [3]. As with other technologies, there are promising new opportunities, but also concerns that specifically affect those involved [4]. One concrete added value that arises from the use of AI is that new frameworks or research tools can be developed that support scientists in evaluating and interpreting large amounts of data. At a conference on the use of AI in higher education, a participant expressed his concerns after the presentation of a prototype that can automatically create interactive, dynamic learning media. I was told that learners' motivation decreases when they know that the learning media used was created by an AI. Whether and how AI-generated learning media influences motivation must be recorded through explicit research. Another concern was that the quality of the learning environment generated by an AI would be worse than that created by a human. The more complex the topic (e.g. computer science), the clearer this should be. In the AI age, however, there is no causality between time and quality. If a person has little time to solve a task, the quality of the solution will suffer. If, on the other hand, they have more time, the probability that the quality of the solution will be better increases. A machine will generate learning media even with little time that is still of high quality. The quality may be influenced by the configuration of the system or model. Since much of the work on AI is anchored in computer science, there is a lack of scientific work on how AI will specifically influence education and thus also teacher training [5].

This article is therefore about capturing current perspectives in order to then make recommendations that can be taken into account in teacher training in order to counteract fears and worries. To achieve this and at the same time to illustrate the added value of AI in the context of statistical analysis, the prototype FACTS-V1 (Filtering and Analysis of Content in Textual Sources) is used. The following chapter describes the method by explaining how the prototype works.

## 2 Methodology

In order to answer the question of how AI will change education, 82 scientific papers from 2024 were analyzed. The papers are available on the peDOCS document server of the Leibniz Institute for Educational Research and Information and can be downloaded without registration. The search term used was "artificial intelligence", which resulted in 500 articles being displayed. FACTS-V1 was used to filter and analyze the content in the text sources. This was created as part of an AI strategy to develop various tools for the education sector. Another tool of this strategy in the field of statistics is ASCVIT, which can be used to automatically record, analyze and interpret numerical and categorical data from data sets [6]. FACTS-V1 consists of a total of three components, the function of which is briefly discussed below.

The first component is a bot that searches the peDOCS platform for scientific articles on a given search term, in this case "artificial intelligence" and year of publication (2024). The articles published in 2024 are downloaded as a PDF file and the exact source information is extracted and saved in a separate text file. The bot's metadata can be customized, allowing the bot and thus the application to be used in any context. The bot is initialized using the Firefox WebDriver (GeckoDriver), which is set up by Selenium. A WebDriver enables the automation of tasks in a browser. It represents a bridge between code and browser. There are different WebDrivers for the different browsers. Selenium is used to automate actions in the browser. It uses the WebDriver to open the website in Firefox. To configure the bot, the peDOCS page was examined using the developer tool (F12) to identify the classes required for navigation. The search query is made according to the defined search term and the results page is loaded. The bot then extracts the links to the relevant articles. The articles are then processed. The subpage of the articles is opened, and the metadata is checked. Further processing only takes place for publications from 2024. The PDF files are saved locally in a defined directory. The bot automatically recognizes when there are additional pages with search results. When the results of one page have been checked, the next page is called up. When there are no more pages, the process is terminated.

The second component of the FACTS-V1 prototype is used to analyze the PDF files collected in the first step. This is achieved by extracting the text, cleaning it and examining it for the defined question using a Large Language Model (LLM), specifically LLama3.1p from Meta. The LLM is installed locally using the Ollama model management system. Communication takes place via the Command Line Interface (CLI). The application first goes through all PDF files. The files are opened and the text is extracted page by page. Unnecessary elements such as page numbers, line breaks and double spaces are removed. The aim of the cleaning is to create a flowing text that is easier for the LLM to read. Since the maximum length of the input context is often limited for LLM, the cleaned text is divided into smaller sections (chunks) of 3500 characters to ensure efficient processing by the model and not to overload it. Each text section is sent to the LLM using a context-related prompt. The model then gives a concise answer to the previously defined question, provided the text section contains relevant information. If the section is not important for the question, a hint (NO ANSWER) is also issued. The results of the analysis are recorded in a text file for each article and section. In addition to the analysis results, files are also created for the cleaned texts. The preparation of the results is clearly structured and therefore easy to understand. The collected data can therefore be quickly used for further analyses.

In the last part of the application, the relevant information from the analyzed text data is used for topic modeling using Latent Dirichlet Allocation (LDA). LDA automatically identifies main topics based on word combinations and their frequency [7]. The extracted relevant answers are saved in a .csv file. The sections that are irrelevant to the question are ignored. The answers are represented by a Bag-of-Words model (BoW), which puts the unstructured text data into a numerical form (vector). The vectors of the BoW model are summarized by the Document Term Matrix (DTM). LDA identifies the main topics, 5 by default, in the texts by analyzing the DTM. The results of the LDA topic modeling are visualized using pyLDAvis. The advantage of this open source library is that the results are quickly understandable and interpretable. The interactive visualization is saved in an HTML file so that it can be opened flexibly in a browser. The functionality of the FACTS-V1 prototype is summarized in the following figure 1.

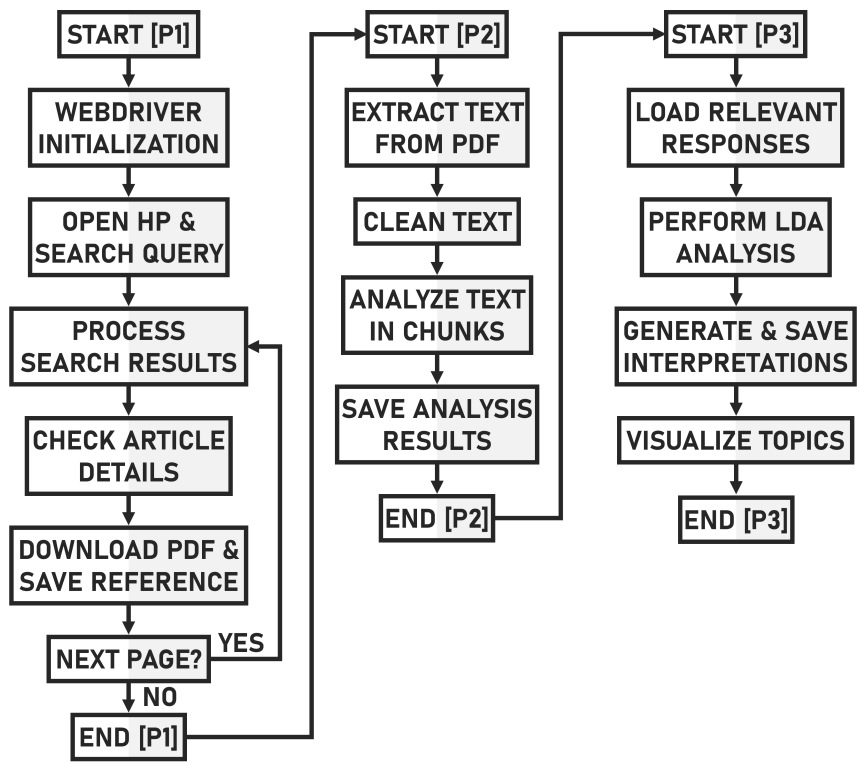

Figure 1: Progression diagram FACTS-V1.

The LDA algorithm was chosen for topic modeling because it is flexible with unknown patterns, is suitable for exploratory analyses and data-driven category formation [7]. The interpretation of the terms of topic modeling can be done in two ways. Either as a manual process based on expert knowledge, intensive text analysis and statistical analysis or by using an LLM. The advantage of an LLM is that the process enables faster, more scalable and more precise interpretations [8]. The manual approach or interpretation by a human, on the other hand, is time-consuming, subjective and limited in terms of scalability, since only samples can be analyzed with large amounts of data [9]. In the FACTS-V1 prototype, the interpretation is carried out by the same LLM (Llama3.1p) that was used to examine the chunks in component 2. The following chapter discusses the results of the LDA topic modeling.

## 3 Results

Of the 82 scientific papers published on the peDOCS document server on the topic of artificial intelligence in 2024, none contained a section that was not relevant to the research question. The full list of articles analyzed can be found in the appendix. Topic modeling with LDA identified five topics characterized by the 30 most common terms within each topic. The clusters show which words frequently appear together in the analyzed papers (see Fig. 2)). The results are based on a relevance value of 1 ($\lambda = 1$), since the aim is to interpret the core terms of the clusters. The global importance of a term in the corpus is measured by the saliency metric [10]. The topic-specific relevance of a term is calculated by the relevance metric [11].

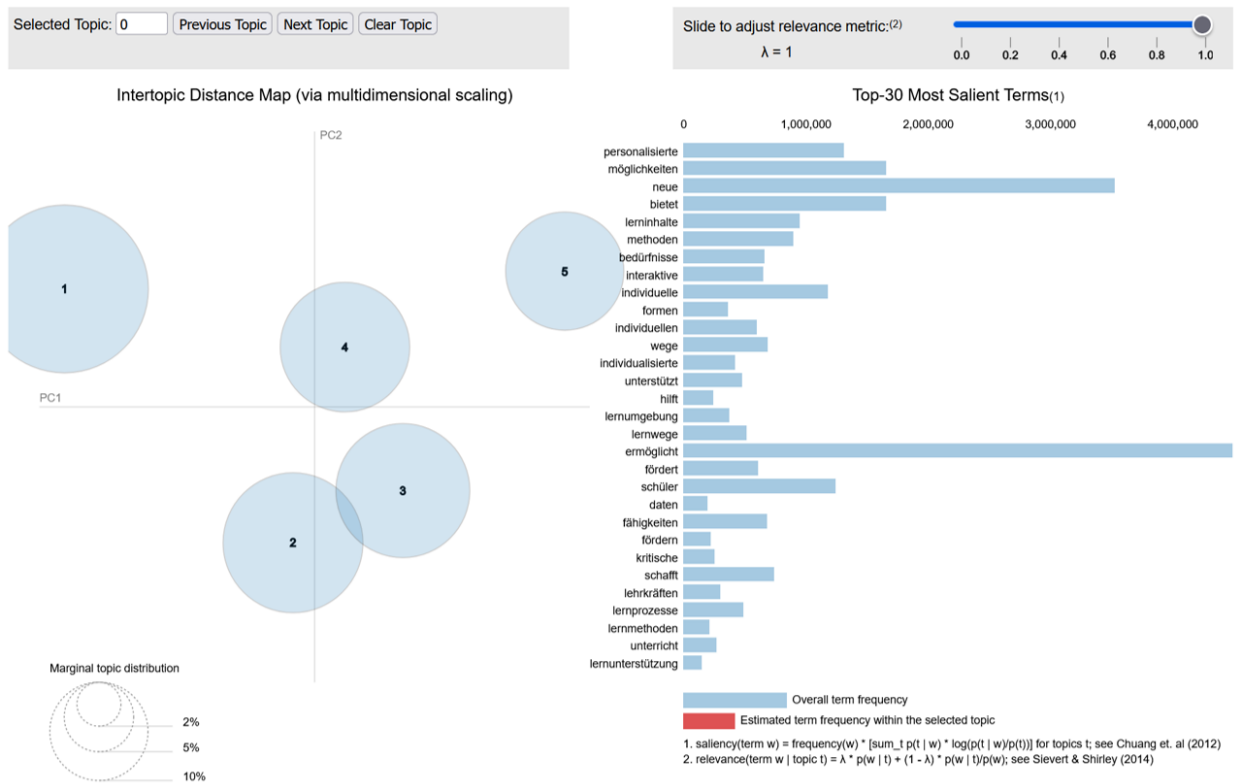

Figure 2: Overview of topic modelling in German (LDA) using pyLDAvis by cluster.

The terms "digital", "promotes", "enables" and "learning" appear in several topics, which indicates overarching key terms. The intertopic distance map shows how similar or different the topics are. There is a slight overlap in clusters 2 and 3. There are topics that are far apart from each other (1 and 5 or 1 and 2 or 3). But there are also topics (2, 3 and 4) that are grouped more closely together. The relative frequency of a topic, i.e. the percentage of the topic in the entire text, is shown by the size of the circle. The figure shows that topic/cluster 1 is the most strongly represented with a share of 29.2 % of the tokens. The second-highest topic weight falls to cluster 2 with a share of 20.3 %. If you click on individual terms, you get more information. For example, you can quickly see that the word "enables" appears most frequently in cluster 1 (see Fig. 3). The blue bar shows the overall frequency of a term throughout the text. When a term is selected, the red bar shows the estimated frequency within the topic. If a term has a long blue bar, it is more likely to be a general term. The Conditional Topic Distribution shows the probability that a term belongs to a specific topic. For example, the term "learning environment" only appears in topic 1.

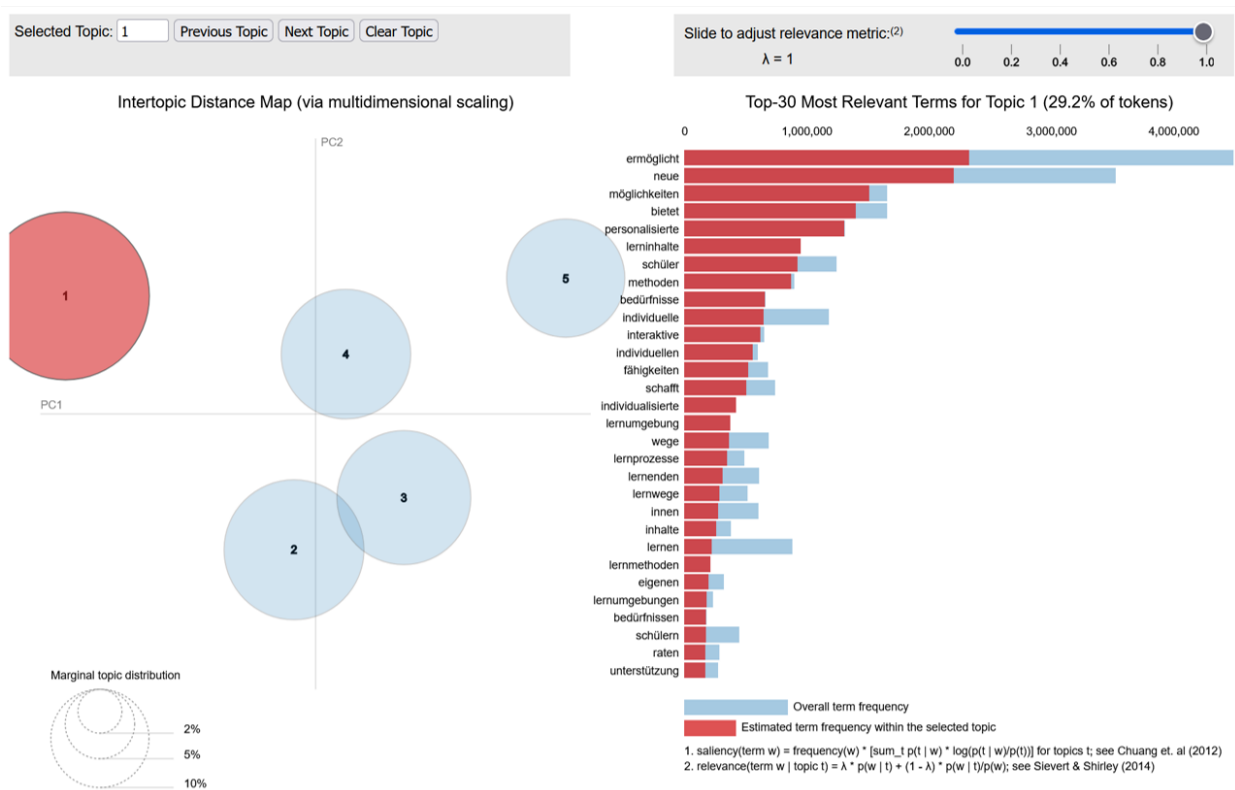

Figure 3: Composition of Cluster in German 1 using pyLDAvis.

The 10 most important terms from clusters 1 to 5 used to interpret the identified themes are as follows (see Table 1):

| Cluster | Terms |
|---|---|
| 1 | individual, needs, methods, students, learning content, personalized, offers, opportunities, new, enables |
| 2 | inside, accompaniment, learning support, digital, promotes, learning, ways, forms, new, enables |
| 3 | as well as, skills, learners, learning, digital, learning paths, promotes, supports, individual, enables |
| 4 | digital, opportunities, skills, teaching, students, promote, offers, enables, learning, new |
| 5 | understanding, critical, develop, critical, promotes, students, data, teachers, enables, helps |

Table 1: Cluster composition (top 10) of topic modeling (LDA).

The restriction to the top 10 terms was chosen in order to achieve a balance between meaningfulness and conciseness. The five clusters and their terms were thematically interpreted in the context of the question "How will the use of AI change education?" by the LLM adapted Llama3.1p as follows (see Table 2):

| Cluster | Topic weight | Theme |
|---|---|---|
| 1 | 29.2% | Individualization of learning |
| 2 | 20.3% | Support and new learning paths |
| 3 | 18.6% | Development of skills |
| 4 | 17.4% | New Possibilities in Teaching |
| 5 | 14.5% | Critical Thinking and Ethical Competence |

Table 2: Thematic interpretation of the clusters by the LLM Llama3.1p.

- The use of AI will enable greater individualization of learning, as machines can respond to the specific needs of learners. Personalized content and customized learning methods can be realized on a broad basis using AI. This will particularly benefit learners with special needs (Cluster 1).

- The use of AI will enable new dimensions of individual support in the learning process. AI can provide learning support and personalized support in a way that opens up both new forms of learning and flexible learning paths. AI makes education more accessible and reduces barriers (Cluster 2).

- The use of AI will promote skills development by using personalized learning paths to use digital tools to enhance learners' skills. The focus here is on promoting digital and technical skills (Cluster 3).

- The use of AI will open up new didactic possibilities in teaching. Teachers will be supported in introducing innovative methods and adapting lessons to the needs of the learners. Specifically, through the use of data-driven teaching strategies and adaptive learning platforms (Cluster 4).

- The use of AI will enable critical thinking and reflection on the use of technology among both learners and teachers. It is important to handle data responsibly and to critically analyze AI systems (Cluster 5).

The results of the LDA topic modeling make it clear that the use of AI will change education in different ways with regard to personalization, competence development, the development of new teaching and learning methods, accessibility, but also with regard to ethical reflection. In the following chapter, the results are interpreted in the context of the research.

## 4    Discussion

The results of the analysis of 82 articles show that there are already concrete ideas in the scientific discussion about how the use of AI will change education. This is made clear by the visualization created by pyLDAvis and the topics found. The topic of "Individualization of learning" (Cluster 1) is the most strongly represented, accounting for almost a third of the entire text corpus (29.2 %), and shows that it is assumed that the use of AI can make learning more individual. The topic of "Support and new learning paths" (Cluster 2) has a topic weight of 20.3 % and shows that AI will play an important role in supporting learning processes. With a topic weight of 18.6 %, the topic of "Development of skills" (Cluster 3) is also important. AI should therefore promote the development of individual and digital skills. The topic of "New possibilities in teaching" (Cluster 4) has a topic weight of 17.4 %. In the articles analyzed, AI is expected to create new opportunities in teaching by integrating digital tools and innovative methods. The topic of "critical thinking and ethical competencies" (cluster 5) has the lowest weight at 14.5 %. The use of AI is intended to promote critical thinking by questioning how this technology works and where the results come from. The result shows that this focus is currently not discussed enough. In the introduction, the concern was described that the use of AI would reduce learners' motivation. The results of the LDA topic modeling show that the term "motivation" does not play a role in any of the identified topics in the context of the question and the articles analyzed. One possible interpretation is that it may not be assumed that the use of AI has an impact on motivation. AI will make education more individual by adapting learning media and learning methods to the needs of the learners. In the future, digital and technological competencies will play a more important role. Teaching and learning processes can be made more innovative through the new opportunities that arise. AI will help to break down barriers in education and make teaching more inclusive. The next step must be to make the opportunities that arise from the use of AI visible through concrete applications. The applications must then be used in practice to answer the following questions. How can AI be used specifically to make education more inclusive and effective? How can ethical questions be resolved when AI is used in education and personal data is collected in the process? If AI is used everywhere, does this lead to dependency? What does the future of education look like in the age of AI? The article has shown that the first version of the FACTS-V1 prototype works. The entire process of data collection, data cleaning, data analysis and data evaluation can be fully automated and can be applied to any context. The application is a concrete example of how the use of AI can establish new research tools and thus support efficiency in educational research. The following chapter summarizes the most important findings and discusses suggestions for future research.

# 5   Conclusion

It is clear that AI will change the field of education. This article uses 82 current scientific papers to show what this change will look like in concrete terms. The heterogeneity of learning groups is still a major problem for teachers. The results of the LDA topic modeling, which was carried out using FACTS-V1, show that scientists expect positive effects from the use of AI, especially in the area of individualization of learning. The use of AI should therefore lead to a fairer educational system by taking differences between learners into account more effectively. For further and deeper analyses, the number of terms per cluster used to interpret a topic could be increased from top 10 to top 20 or 30. However, the loss of relevance must be taken into account here, as lower probabilities or relevance values contribute less to characterizing a topic [11]. To identify unique terms or distinguishing features specific to a topic, the relevance value could be set from 1 to 0, which would display the terms with the highest log-lift value. Alternatively, the default value of 0.6 could be used to combine both approaches (probability/lift-weighted terms). To check how the answer to a question changes over the years, a similar analysis could be carried out with a different publication year and the results of the topic modeling could be compared. In further analyses in other disciplines, the time required could be recorded to document how long certain steps, specifically the text analysis by the LLM, take. Here, different models could be compared to check how the performance of the application improves and whether certain models produce better results. The relevant answers collected in the second step of the prototype could also be statistically evaluated by another method. Automatic clustering using the K-Means algorithm would also be possible. Rule-based categorization, in which keywords are manually defined for each category, is also possible [12]. Clear categories would then already have to be in place. The results of the topic modeling could also be analyzed in more detail. Specifically by identifying frequently occurring n-grams in order to identify trends if necessary. Deeper semantic analyses are also possible by using Named Entity Recognition (NER) to record the topics, actors, technology or organizations [13]. Network diagrams could be used to visualize possible relationships between keywords such as AI and learning content. The examples mentioned illustrate the numerous possibilities that arise from the results of the FACTS-V1 prototype. These findings offer a promising outlook on future developments in the field of education, which can be further advanced through the use of AI.

# A   Appendix

The following sources were used as data basis for the LDA topic modeling and are available on peDOCS via open access.